\documentclass[conference]{IEEEtran}
\IEEEoverridecommandlockouts
\usepackage{cite}
\usepackage{makecell}
\usepackage{amsmath,amssymb,amsfonts}
\usepackage{algorithmic}
\usepackage{graphicx}
\usepackage{subfigure}
\usepackage{multirow}
\usepackage{array}
\usepackage{threeparttable} 
\usepackage{tablefootnote} 

\usepackage{textcomp}
\usepackage{xcolor}
\def\BibTeX{{\rm B\kern-.05em{\sc i\kern-.025em b}\kern-.08em
    T\kern-.1667em\lower.7ex\hbox{E}\kern-.125emX}}

\begin{document}

\title{A Memory-Efficient Retrieval Architecture for RAG-Enabled Wearable Medical LLMs-Agents}

\author{\IEEEauthorblockN{Zhipeng Liao\textsuperscript{1,*}, Kunming Shao\textsuperscript{2,*}, Jiangnan Yu\textsuperscript{2},  Liang Zhao\textsuperscript{3}, \\Tim Kwang-Ting Cheng\textsuperscript{2}, Chi-Ying Tsui\textsuperscript{2}, Jie Yang\textsuperscript{1,4,†}, Mohamad Sawan\textsuperscript{1,4,†}}

 \thanks{*Both authors contributed equally.}


 \IEEEauthorblockA{\textsuperscript{1}CenBRAIN, Westlake University, Hangzhou, China, \\
 \textsuperscript{2}The Hong Kong University of Science and Technology, Hong Kong SAR, China, \\
 \textsuperscript{3}South China University of Technology, Guangzhou, China\\
 \textsuperscript{4}Integrated-On-Chips Brain-Computer Interfaces\\ Zhejiang Engineering Research Center, Hangzhou, China
 }
  \vspace{-8mm}

}

\maketitle

\begin{abstract}

With powerful and integrative large language models (LLMs), medical AI agents have demonstrated unique advantages in providing personalized medical consultations, continuous health monitoring, and precise treatment plans. Retrieval-Augmented Generation (RAG) integrates personal medical documents into LLMs by an external retrievable database to address the costly retraining or fine-tuning issues in deploying customized agents. While deploying medical agents in edge devices ensures privacy protection, RAG implementations impose substantial memory access and energy consumption during the retrieval stage. This paper presents a hierarchical retrieval architecture for edge RAG, leveraging a two-stage retrieval scheme that combines approximate retrieval for candidate set generation, followed by high-precision retrieval on pre-selected document embeddings. The proposed architecture significantly reduces energy consumption and external memory access while maintaining retrieval accuracy. Simulation results show that, under TSMC 28 nm technology, the proposed hierarchical retrieval architecture has reduced the overall memory access by nearly 50$\%$ and the computation by 75$\%$ compared to pure INT8 retrieval, and the total energy consumption for 1 MB data retrieval is 177.76 $\mu$J/query.

\end{abstract}






\section{Introduction}

AI agents utilizing Large Language Models (LLMs) have demonstrated significant effectiveness in various applications, including analysis, comprehension, decision-making, etc.~\cite{ClinicalBERT,FinBERT,LEGAL-BERT}. Recent works have also shown their promising ability to offer personalized medical services~\cite{li2024agent, calisto2022breastscreening, vicari2003multi, schmidgall2024agentclinic}. However, when deploying AI agents in medical scenarios, substantial volumes of sensitive data—including personal health metrics, medical records, and physiological indicators—present critical privacy challenges for medical AI agent implementations. Transmitting private data to cloud infrastructure or using it in public LLM training datasets poses significant privacy risks. Moreover, the resource requirements for LLM fine-tuning on edge devices to integrate private data remain prohibitively expensive~\cite{shen2024towards, lyu2024rethinking}.

\begin{figure}[t]
\centering
\includegraphics[width=0.85\columnwidth ]{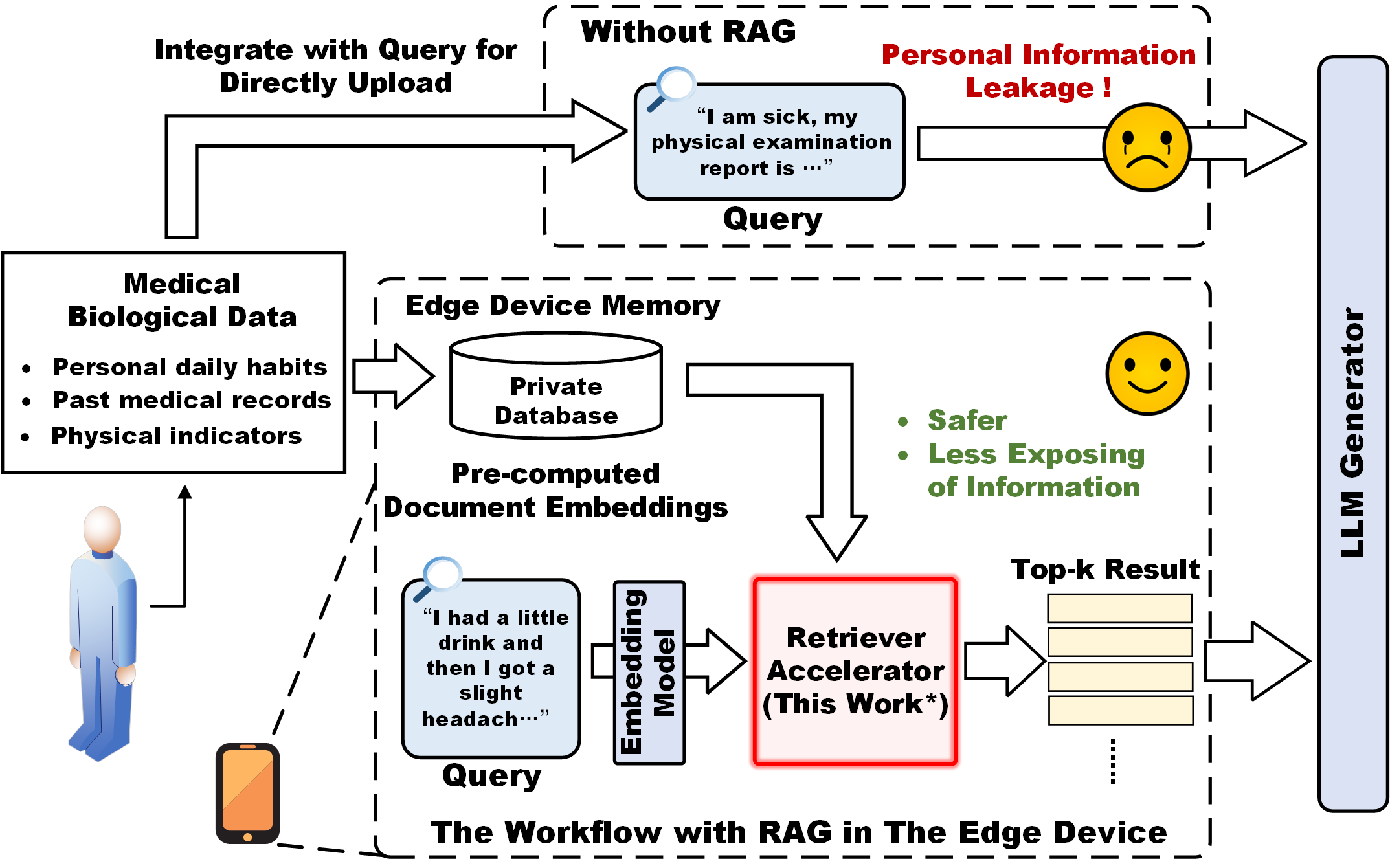}
 \caption{RAG-enabled medical LLM agent on edge wearable device }
 \label{introduction}
 \vspace{-6mm}
\end{figure}
\begin{figure*}[t]
\centering
\includegraphics[width=0.9\textwidth]{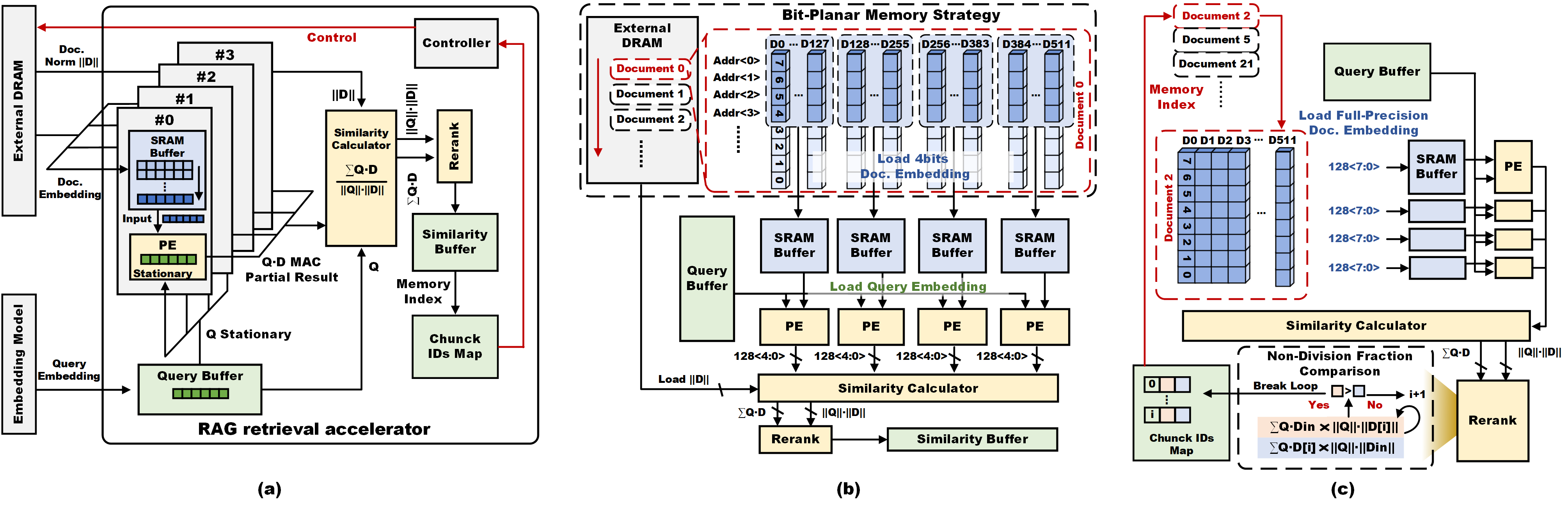}
 \caption{(a) Architecture of RAG retrieval accelerator with query stationary dataflow, (b) Stage 1: MSB INT4 approximate retrieval workflow and bit-planar memory strategy, (c) Stage 2: INT8 full-precision retrieval workflow and non-division fraction comparison} 
 \label{relatedwork}
 \vspace{-4mm}
\end{figure*}

The Retrieval-Augmented Generation (RAG) technique provides an efficient and effective solution by integrating private medical information, such as health records, diagnostic imaging, and genomic data, as a searchable external knowledge database for LLMs~\cite{lewis2020retrieval,zhao2024retrieval,yan2024corrective,ren2024retrieval,tian2025npu}. As shown in Fig. 1, personal health data such as medical records and physiological indicators are converted into document embeddings through an embedding model during the offline phase and then stored in the memory of edge wearable devices. User interactions are encoded into a high-dimensional query embedding vector through the embedding model, facilitating similarity computations to compare against the stored document embeddings. By integrating the retrieved relevant document chunks with the original query as an augmented prompt, the LLM’s capability can be enhanced to generate more accurate and practical content based on relevant personal data. The RAG technique ensures that private data does not need to be made public or included in the cloud-based training of LLMs, thereby safeguarding data privacy and security. Furthermore, when handling diverse medical information from different individuals, LLMs can offer more personalized recommendations without the need for retraining or fine-tuning the model. 

However, the retrieval process for large-scale document embeddings presents significant challenges on external memory access, energy consumption, and latency. The limited capacity of on-chip SRAM results in intensive data transfers between off-chip memory and processing units, causing substantial latency and energy overhead, and ultimately making the retrieval process a bottleneck in RAG systems \cite{shen2024towards,quinn2025accelerating,jo202523,fang2023510,tian2023neurocare,zhao2022emerging,fang2024energy}.


This paper proposes an efficient edge RAG retrieval architecture for resource-constrained wearable systems, featuring three key innovations: 
\subsubsection{\textbf{Quantization-Aware Two-Stage Hierarchical Retrieval: }}The architecture employs a two-stage retrieval scheme, first reading only the most significant 4 bits of INT8 embeddings to generate an approximate candidate set, then performing full 8-bit precision retrieval exclusively on this candidate set. 
\subsubsection{\textbf{Bit-Planar Storage Strategy: }}This paper implements a bit-planar storage approach, storing 512-dimensional embeddings in DRAM with a 512-bit width, where each DRAM row stores one bit of INT8 data. This enables selective access to the most significant 4 bits for approximate retrieval while keeping the lower 4 bits in DRAM without moving them. 
\subsubsection{\textbf{Query Stationary Dataflow: }}This paper implements a query stationary dataflow by maintaining the query embedding vector static on the 1-D Processing Element (PE) array while sequentially loading document embeddings into the PE, thereby improving PE utilization and data reuse.

\section{The proposed architecture}

\subsection{Overall Hardware Architecture}

As depicted in Fig. 1, medical information is transformed into high-dimensional embedding vectors using a transformer-based embedding model~\cite{reimers2019sentence}. In this paper, each document chunk is represented by a 512-dimensional embedding vector, with each entry quantized and stored in INT8 format. During retrieval, commonly used methods include cosine similarity and Maximum Inner Product Search (MIPS), as illustrated by the following equations:
\[
Query = [q_1, q_2, \ldots, q_n], \quad Doc = [d_1, d_2, \ldots, d_n]
\]

\vspace{-3mm}
\[
Cos\_Sim(Query, Doc) = \frac{\sum_{i=1}^{n} q_i d_i}{\sqrt{\sum_{i=1}^{n} q_i^2} \cdot \sqrt{\sum_{i=1}^{n} d_i^2}}
\]
\[
MIPS(Query, Doc) = \sum_{i=1}^{n} q_i d_i
\]
\vspace{-1mm}

Both cosine similarity and MIPS are based on Multiply-Accumulate (MAC) operations, which can be processed in parallel. As shown in Fig. 2(a), we propose a hardware architecture to support and accelerate similarity calculations, including MAC, norm computation, similarity calculation, and reranking operations. Our architecture supports both cosine similarity and MIPS. The choice of similarity retrieval method depends on the type of embedding model: If the embeddings are normalized, cosine similarity is preferred for better accuracy; If the embeddings are unnormalized, MIPS using pure MAC operations is adopted~\cite{mussmann2016learning}.

During runtime, the medical or life-related query is encoded into a 512-dimensional query embedding vector, which is stored in the query buffer before being loaded into the PE and kept stationary. Once the query is ready, document embeddings are sequentially loaded from off-chip DRAM for similarity computation. Due to limited on-chip resources, it is not feasible to cache all document embeddings in the SRAM buffer. Therefore, we employ four small dual-port SRAM buffers for streaming, each 128 bits wide and responsible for storing 128 dimensions of the document embedding. The loaded 512-dimensional embedding is divided into four parts to alleviate on-chip bandwidth congestion of the SRAM buffer.

Four PEs are placed near the SRAM to execute MAC operations. Each PE is responsible for 128-dimensional INT4 MAC operations, and the calculated partial sums are fused using the similarity calculator. The similarity calculator unit is also used for query vector norm calculation and the final cosine similarity computation. Taking cosine similarity as an example: the query norm is calculated via the similarity calculator during query loading, while the document norm is pre-calculated and stored in DRAM. 

With the MAC result, query norm, and document norm, the cosine similarity can be computed and ranked using non-division fraction comparison, as shown in Fig. 2(c). After similarity calculation and reranking, the document chunk IDs with the highest similarity values are recorded in the chunk IDs map.


\subsection{Stage 1: MSB INT4 Approximate Retrieval}

As shown in Fig. 2(b), in the first computational stage, we employ an approximate retrieval scheme based on the MSB INT4 of the document embedding. In this stage, a dataflow pipeline is constructed to generate a candidate set of approximately the top-50 results. First, the embedding data of multiple document chunks is stored in different blocks of off-chip memory, with each block of 8 rows storing the embedding vector for one document chunk. The DRAM width matches the dimension of the embedding vector, enabling the use of the bit-planar memory strategy to store 512 INT8 data in 8 rows, with each row storing one bit of all 512 INT8 data. 

The Bit-Planar Memory Strategy enables bit-wise reading of the embedding vector, allowing the upper 4 bits of the INT8 data to be accessed without reading the unnecessary lower 4 bits. This approach reduces DRAM access latency and energy consumption by 50\%, which are the dominant contributors to overall system latency and power consumption\cite{ko2025cosmos,shen2024towards}.

\begin{figure}[t]
\centering
\includegraphics[width=0.65\columnwidth ]{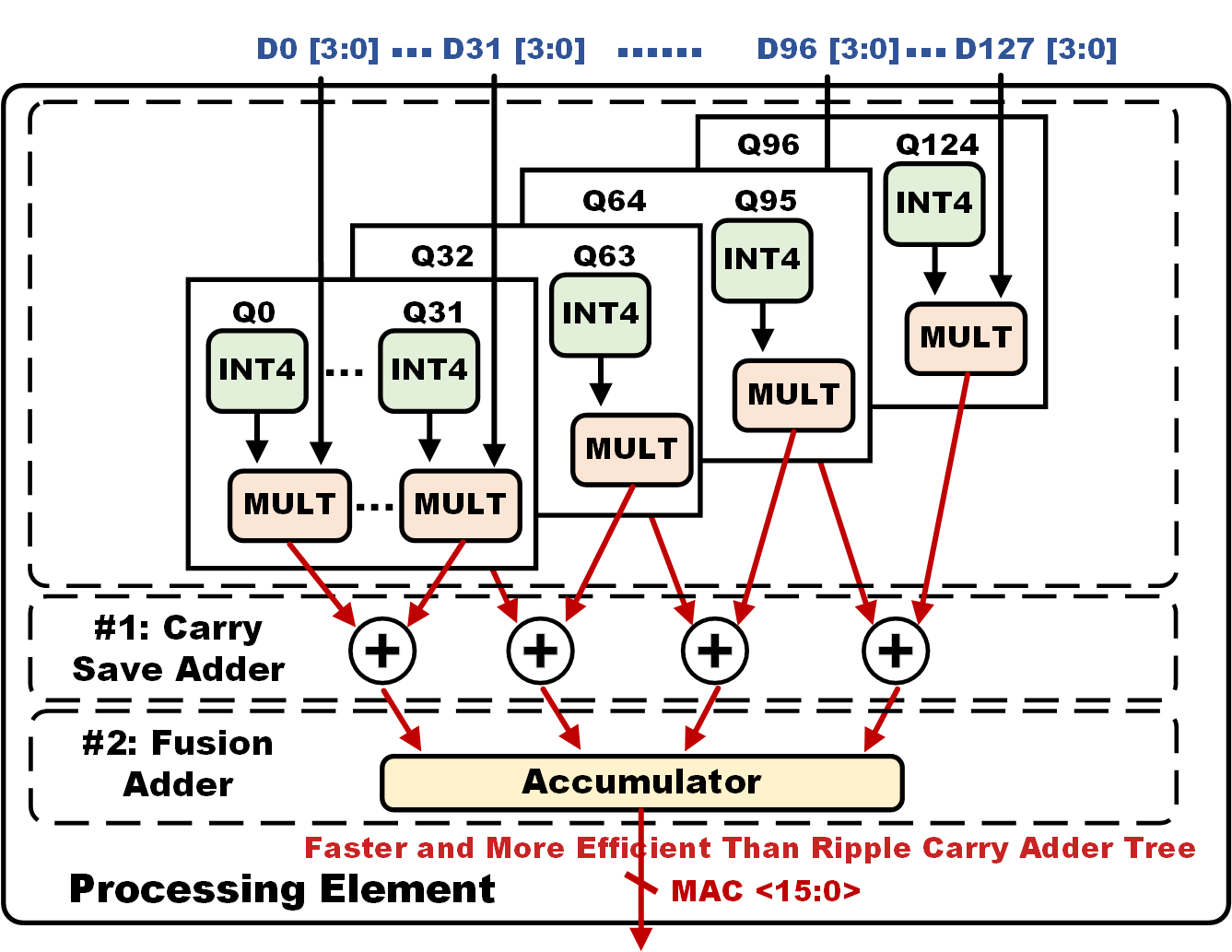}
 \caption{ Structure of the Processing Element (PE)}
 \vspace{-4mm}
\end{figure}
Subsequently, the data is sent to the PE to compute the MAC result of the two embeddings, as illustrated in Fig. 3. Each PE processes 128 INT4 data elements, totaling 512 bits. To avoid a long combinational critical path during multiplication and accumulation of large-scale data, the additions are performed using a pipelined two-stage carry-save adder structure after the 4-bit multiplications of corresponding entries\cite{zhao2025flexible, shao2024syndcim}. In the first stage, 64 dimensions are processed in parallel by multiple carry-save adders, while in the second stage, a fusion adder further accumulates the results.

Finally, the similarity calculator performs further fusion across PEs and outputs the MAC results of the query and document embeddings, along with the query norm and document norm, to the rerank module. Since there is no significant performance difference between INT4 and INT8, and typically only a limited number of document chunks are needed (e.g., top-1, top-3, top-5), we select the top-50 as the size of our candidate set. Through dense comparison, the module ranks the similarities from highest to lowest and generates a candidate set containing the top-50 document chunks.

\subsection{Stage 2: INT8 Full Precision Retrieval}
As depicted in Fig. 2(c), after the approximate retrieval stage, this architecture will switch to a high-precision refinement stage. During the full precision retrieval stage, the previously generated top-50 candidate documents in the first stage will be reloaded and re-retrieved using INT8 full precision. The chunk IDs map clearly sorts the document IDs and corresponding root addresses of the approximate top-50 chunks, such as Document 2, Document 5, Document 21, etc. Subsequently, corresponding top-50 INT8 document embeddings will be loaded to the on-chip SRAM for a new stage retrieval.

This quantization-aware hierarchical retrieval approach can significantly reduce external DRAM access and computational intensity while maintaining comparable retrieval accuracy.

\section{Experimental and Validation Results}
\subsection{Experimental Setup}

For the hardware experiment, we implemented an RAG retrieval accelerator for edge wearable devices based on the TSMC 28 nm process. The SRAM buffer of this accelerator was generated using the TSMC 28 nm SRAM compiler, while the remaining digital circuits were synthesized using Synopsys Design Compiler and then automatically placed and routed on the Cadence Innovus platform. Finally, post-layout simulation was performed to evaluate the overall latency and energy consumption on Synopsys VCS and PrimeTime.

In the software experiment, we utilized the open-source BEIR framework \cite{thakur2021beir} and evaluated the retrieval precision of INT4 and INT8 quantized embeddings \cite{jacob2018quantization,kunming2025dircrag} and the proposed hierarchical retrieval in three datasets of different domains, including SciFact \cite{wadden2020fact}, NFCorpus \cite{boteva2016full}, and ArguAna \cite{wachsmuth2018retrieval}. The embedding model used in this system is the complete MiniLM-L6-v2 \cite{wang2020minilm}, and it is combined with the open-source Sentence-BERT \cite{reimers2019sentence} framework, with an embedding vector dimension of 512. The retrieval performance of RAG is measured by Precision@k (P@k), which reflects the proportion of relevant document chunks to the query among the retrieved top-k results.

For the simulation of the complete system, we developed a Python simulator to evaluate the energy efficiency and retrieval delay of the architecture proposed in this paper. The simulator simulated and analyzed the energy consumption, interconnection bandwidth, and cycle delay of hardware components such as PEs, similarity calculator modules, SRAM buffers, global top-k rerank, etc. 

\begin{figure}[!t]
\centering
\includegraphics[width=0.90\columnwidth ]{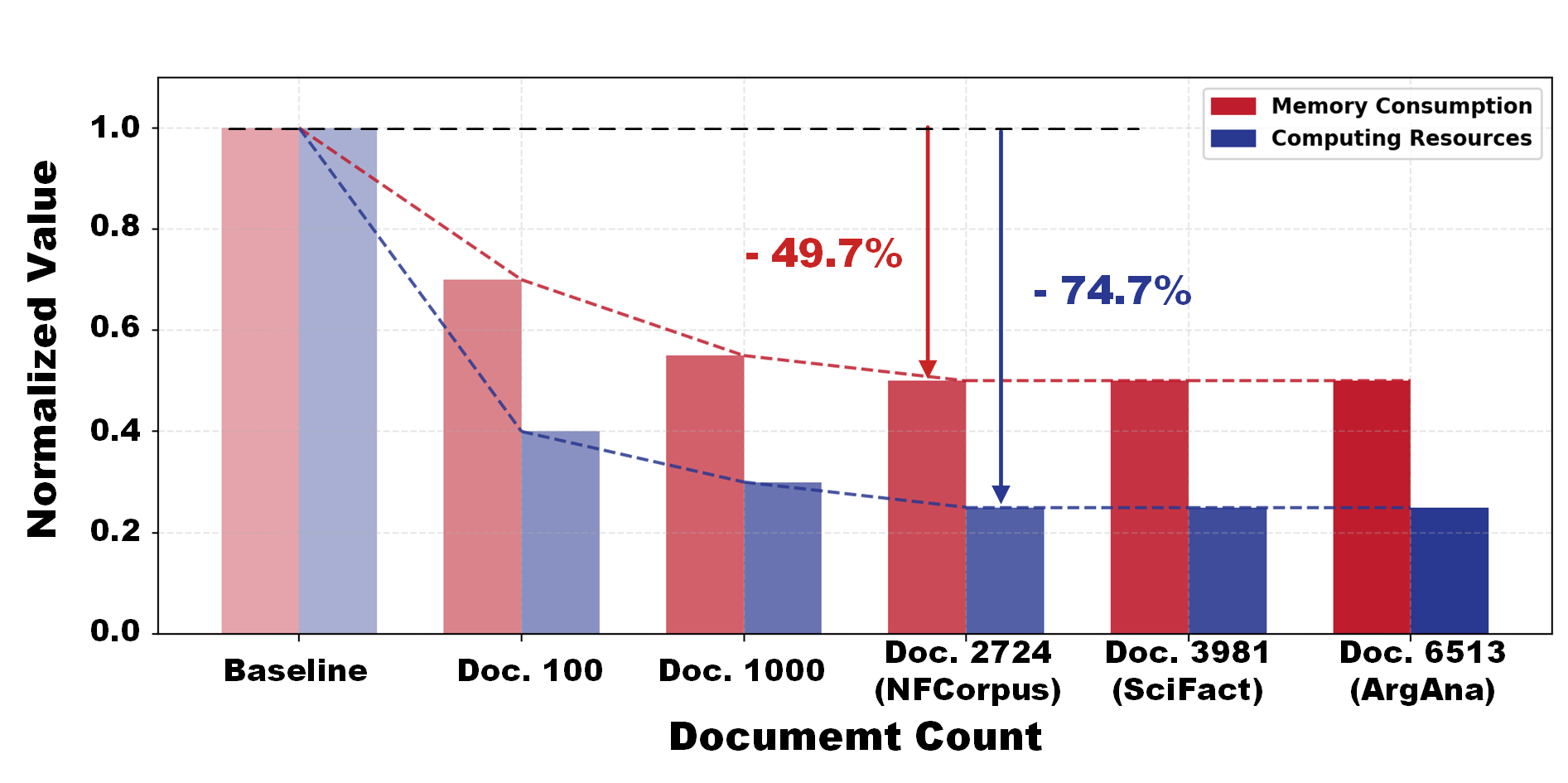}
 \caption{Memory access reduction and computation reduction for different document chunk numbers. }
 \vspace{-4mm}
\end{figure}

\subsection{Software Evaluation}
This study compared the performance differences between the traditional retrieval method and the proposed two-stage retrieval scheme in terms of memory usage and computational load. Fig. 4 shows the acceleration effectiveness for reducing external memory access and computation under different amounts of document chunks. As the number of chunks increased from 100 to nearly 10000, the proposed two-stage retrieval scheme reduced memory utilization load from 30$\% $ to nearly 50$\%$ and decreased computational load from 55$\% $ to 74.7$\% $. Therefore, the proposed hierarchical two-stage retrieval reduces the memory usage and computational cost by nearly 50$\%$ and 75$\%$ respectively for actual datasets. It can be concluded that the quantization-aware two-stage retrieval scheme proposed in this paper outperforms the traditional method in terms of memory access and computational load. It indicates that this scheme has better resource utilization efficiency when dealing with large-scale document sets.

For the accuracy simulation, we conducted two baseline experiments using pure INT4 and INT8 quantization, respectively. Table I compares the retrieval precision@1 (P@1) of the proposed two-stage retrieval scheme with those of the baseline INT8 and INT4 data formats across multiple datasets. As shown in Fig. 5(a) and Table I, the two-stage retrieval achieves P@1 values on all three datasets that are very close to those of pure INT8, whereas pure INT4 exhibits a more noticeable drop in precision. Furthermore, as illustrated in Fig. 5(b), our simulations demonstrate that the two-stage retrieval scheme can significantly reduce energy consumption, bringing it down to the level of pure INT4, while still maintaining retrieval precision comparable to pure INT8.
In summary, the two-stage retrieval scheme effectively reduces external memory access and computational load while maintaining a comparable retrieval precision.

 \begin{figure}[!t]
\centering
\includegraphics[width=0.95\columnwidth ]{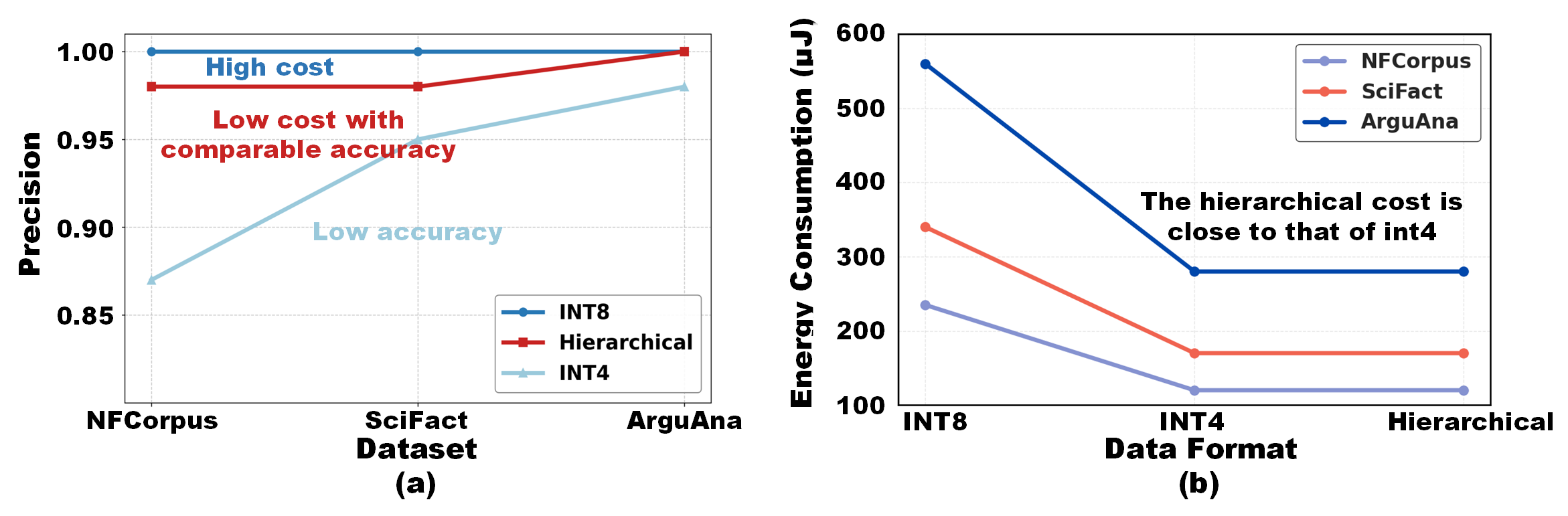}
 \caption{Data format of the retrieval method proposed in this paper - hierarchical - is compared with other data formats. (a) Normalized retrieval precisions. (b) Energy consumption per query}
 \vspace{-4mm}
\end{figure}

\subsection{Hardware Evaluation}
In this study, we designed an RAG retrieval accelerator architecture for wearable edge devices and evaluated it using post-layout simulations, as summarized in Table II. The DRAM energy data are sourced from \cite{horowitz20141,sze2017efficient}, while all other results are derived from our own simulations. For a 1MB INT8-quantized document embedding database with 512 dimensions, our simulation results show that querying the entire database consumes only 178 $\mu$J. Furthermore, due to the limited on-chip reuse of document embeddings, 99\% of the total energy consumption is attributed to off-chip data transfers, whereas on-chip processing logic accounts for just 0.2\% of the system’s overall energy usage. Therefore, the main advantage of the proposed two-stage retrieval scheme is its ability to reduce data transfer from off-chip DRAM to on-chip SRAM, thereby significantly lowering DRAM-related power consumption.


According to the results in Table III, when retrieving the same SciFact dataset, our accelerator achieves comparable precision@1 to NVIDIA's flagship RTX3090 GPU - a standard benchmark in AI/ML workloads. While the RTX3090 operates on 8 nm process with 628.4 mm² die size, our 28 nm accelerator maintains similar functionality in just 0.077 mm² area while consuming nearly two orders of magnitude less energy. Designed for edge devices, this accelerator features a highly compact area and ultra-low energy consumption, yet delivers comparable functionality and retrieval efficiency.

\begin{table}[!t]
\centering
\caption{ Comparison of retrieval quality under different data accuracies for different data sets}
\label{tab:energy_consumption}

\resizebox{0.65\columnwidth}{!}{
\begin{tabular}{|l|c|c|c|}
\hline
P@1& INT8&  INT4
&Hierarchical (This work)\\
\hline
NFCorpus& 0.421&  0.368
&0.412\\
\hline
SciFact& 0.507&  0.483
&0.497\\
\hline
ArguAna& 0.253&  0.248&0.253\\\hline
\end{tabular}

}
 \vspace{-3mm}
\end{table}
\begin{table}[!t]
\centering
\caption{Energy consumption analysis of different modules}
\label{tab:energy_consumption}
\begin{threeparttable}
\resizebox{0.7\columnwidth}{!}{
\begin{tabular}{|l|c|c|c|}
\hline
Module & pJ/bit &Energy/Query* &Proportion\\
\hline
DRAM & 40 &176 $\mu$J&98.831$\%$\\
\hline
SRAM & 0.2 &1.72 $\mu$J&0.966$\%$\\
\hline
PE & 0.0078&343.5 nJ&0.193$\%$\\
\hline
Similarity Calculator& 0.0003&13.6 nJ&0.008$\%$\\
\hline
Rerank& 0.0001&5.5 nJ&0.003$\%$\\
\hline
\end{tabular}

}
\begin{tablenotes}
\item[$\ast$] For 1MB INT8 document embeddings. 
\end{tablenotes}
\end{threeparttable}
\vspace{-3mm}
\end{table}
\begin{table}[!t]
\centering 
\caption{Comparison between our accelerator and other works}
\begin{threeparttable}

\resizebox{0.95\columnwidth}{!}{ 
\begin{tabular}{|c|c|c|c|} \hline
Work                                                                   &  RTX3090
 &1FPGA+2GPU\cite{jiang2023chameleon}&This work*                                 \\ \hline
Tech (nm)&  Samsung 8 nm&\makecell{16 nm (FPGA),  8 nm (GPU)}&TSMC 28 nm\\ \hline
Freq (MHz)&  1395
 &\makecell{140 (FPGA) \\1395 (GPU)}&200$\sim$400                                                                                                                                           \\ \hline
Area (mm$^{2}$)&  628.4
 &1256.8+FPGA&0.077 \\ \hline
Embeddings&  FP32
 &FP32&INT4/8 \\ \hline
Energy/Query&                                                                                                                                              86.8 mJ&95.6 mJ&337.74 $\mu$J\\\hline
SciFact P@1&  0.507 &-&0.497\\\hline
\end{tabular}}
\begin{tablenotes}
\item[$\ast$] This work uses post-layout estimation.
\end{tablenotes}
\end{threeparttable}
\label{table3}
\vspace{-5mm}
\end{table}

\section{Conclusion}
In conclusion, this paper presents a hierarchical retrieval architecture for edge RAG applications. The proposed retrieval process consists of two stages: approximate retrieval followed by full-precision retrieval. This approach significantly reduces both data transmission and memory usage for off-chip memory and SRAM components, resulting in exceptionally low retrieval energy consumption. Extensive simulation results rigorously demonstrate the effectiveness of this acceleration architecture, showing that query energy consumption on the SciFact dataset is only 337.74 $\mu$J, with only a minimal decrease in retrieval accuracy. These promising results indicate that the proposed architecture is a practical and efficient solution for high-performance RAG tasks on resource-constrained edge devices.

\section{Acknowledgement }
This work was supported in part by the STI2030-Major Projects under Grant 2022ZD0208805, in part by the “Pioneer” and “Leading Goose” Research and Development Program of Zhejiang under Grant 2024C03002, and in part by the Key Project of Westlake Institute for Optoelectronics under Grant 2023GD004. (Corresponding authors: Jie Yang; Mohamad Sawan.)

\clearpage

\newpage
\bibliographystyle{IEEEtran} 
\bibliography{IEEEabrv,reference}

\end{document}